\documentclass[12pt]{article}
\usepackage{cite}
\usepackage{pifont,authblk}
\usepackage{amssymb,amsbsy,amsmath,amscd,amsthm}
\usepackage{tikz,pgfplots}
\usetikzlibrary{calc,fadings,decorations.pathreplacing,decorations}
\newtheorem{defn}{Definition}[section]
\newtheorem{theorem}{Theorem}[section]
\newtheorem{lemma}{Lemma}[section]
\newtheorem{hypo}{Hypothesis}[section]
\newcommand\pgfmathsinandcos[3]{%
  \pgfmathsetmacro#1{sin(#3)}%
  \pgfmathsetmacro#2{cos(#3)}%
}
\newcommand\LongitudePlane[3][current plane]{%
  \pgfmathsinandcos\sinEl\cosEl{#2} 
  \pgfmathsinandcos\sint\cost{#3} 
  \tikzset{#1/.estyle={cm={\cost,\sint*\sinEl,0,\cosEl,(0,0)}}}
}

\newcommand\DrawLongitudeCircle[2][1]{
  \LongitudePlane{\angEl}{#2}
  \tikzset{current plane/.prefix style={scale=#1}}
  \pgfmathsetmacro\angVis{atan(sin(#2)*cos(\angEl)/sin(\angEl))} %
  \draw[current plane] (\angVis:1) arc (\angVis:\angVis+180:1);
  \draw[current plane,dashed] (\angVis-180:1) arc (\angVis-180:\angVis:1);
}

\begin{tikzfadingfrompicture}[name=custom fade]%
\path(-0.2cm,0.2cm) rectangle (1.2cm,-2cm); 
\pgfinterruptboundingbox
\draw[very thick,transparent!20,->] (0cm,0cm) .. controls +(0cm,-1cm) and
+(0cm,1cm) .. (1cm,-2cm);
\endpgfinterruptboundingbox
\end{tikzfadingfrompicture}
\renewcommand{\omega}{\varpi}
\newcommand\email[2]{#1@#2}

\def\x{{x}}

\def\B{\mathcal{B}}
\def\N{\hat{N}}

\def\C{{\mathbf C}}
\def\R{{\mathbf{R}}}

\def\Z{{\mathbf{Z}}}
\newcommand\torus[1]{{T}^{#1}}
\def\T{{\mathcal{T}}}
\def\trgr{\mathfrak{T}}
\def\G{{\Lambda}}

\def\eq#1{(\ref{#1})}

\def\rt{\longrightarrow}
\renewcommand\hat{\widehat}

\newcommand\ie{\emph{i.e.~}}

\def\map{\longmapsto}
\newcommand\ind{\operatorname{ind}}
\newcommand\coker{\operatorname{coker}}
\newcommand\ch[1]{\operatorname{ch}({#1})}
\newcommand\ket[1]{|#1\rangle}
\def\SO{\operatorname{SO}}
\def\SU{\operatorname{SU}}
\def\sph{\operatorname{S}}
\begin{document}
\title {Topology behind topological insulators}
\author[1]{Koushik Ray \thanks{\email{koushik}{iacs.res.in}}}
\affil[1]{Indian Association for the Cultivation of Science \\
Calcutta 700~032. India.}
\author[2]{Siddhartha Sen  
\thanks{\email{siddhartha.sen}{tcd.ie}, \email{sen1941}{gmail.com}}}
\affil[2]{CRANN \& School of Physics, Trinity College Dublin, Dublin--2, 
Ireland}
\date{}
\maketitle
\vfil
\begin{abstract}
\noindent
In this paper topological $K$-group calculations for fiber bundles with structure group $\SO(3)$ over tori are carried out to explain why topological insulators have special conducting points on their surface but are bulk insulators. It is shown that these special points are gap-less and conducting for topological reasons and follow from the $K$-group calculations. The existence of gap-less surface points is established with the help of an additional topological property  of the $K$-groups which relates them to the index theorem of an operator. The index theorem relates zeros of operators to topology. For the topological insulator the relevant operator is a Dirac operator, that emerges in the problem because  the system has strong spin-orbit interactions and time-reversal invariance.  Calculating $K$-groups over tori require some special topological tools that are are not widely known. These are explained. We then show that the actual calculation of $K$-groups over tori becomes straightforward once a few topological results are in place. Since  condensed matter systems with periodic lattices, are always bundles over tori the procedures described is of general interest.
\noindent
\end{abstract}
\thispagestyle{empty}
\section{Introduction}
\label{sec:intro}
Topological methods are increasingly being used in
different branches of physics \cite{Arnold,Monastyrsky,Bhattacharjee}. 
A striking example of the usefulness of  topology in physics was the prediction 
\cite{Zhang1,Zhang2}
that a bulk insulator of a certain kind could have special 
surface points that are conducting. It was also realized that a proper
topological understanding of this observational result required
a $K$-group calculation \cite{fu,kauf}.  However such a calculation presented
in a way that it can be readily used is not available.
In this paper we carry out such a calculation which requires calculating the topological 
$K$-groups of bundles over tori. The essential point is that the non 
vanishing of $K$-groups can explain the existence of surface gap-less conducting points. This is  the topological reason behind the topological insulator.  In order to carry out the calculation
 we have to explain a number of the topological notions and also use 
the fact that the system has strong  spin-orbit interactions 
and time-reversal symmetry. These requirements lead to the emergence of an effective relativistic Dirac like equation at special points.   
 
Since the topological tools required to calculate the $K$-groups are
not widely known we provide the details required for the calculation in an intuitive way without proving results stated. Once a few key topological results are in place,
the actual calculation of $K$-groups for bundles over tori is straightforward.
Since many condensed matter systems are bundles over tori the 
methods described should be of general interest. 

Topology studies features of a system that are invariant under continuous deformations. Two
spaces $X,Y$ are topologically equivalent (homeomorphic) if there is a one-to-one invertible map between them. Unfortunately there is no method available for establishing such an equivalence for arbitrary topological spaces $X,Y$.  There is a weaker form of topological equivalence called homotopy equivalence. Now two spaces $(X,Y)$ are homotopy equivalent
if one can be converted to the other by a continuous map. The 
$n$-dimensional Euclidean $\R^n$ is thus homotopy equivalent to a point since we have the continuous map $\R^n\rightarrow tx$ with $x$ an arbitrary point of $\R^n$ and $t$, the homotopy parameter where $0\leq t \leq 1)$ converts all point $x$ to the origin of $\R^n$. Our discussions will use homotopy equivalence to describe topological properties.

The basic building blocks of topological spaces are open sets (or closed sets) since these objects do not change their character under continuous deformations. Arbitrary spaces $X$ can be regarded as a topological space if it can be described in terms of open (or closed) sets in the following way. We introduce a collection of open sets associated with $X$ so that  includes $X$ itself and the null set $\phi$. It is then required that the collection of these open sets under arbitrary union and under finite intersection are open sets. There are many ways of choosing these open sets. Each way defines a topology for $X$. For example the surface of a sphere can be described using two open sets one covering the northern and the other covering the southern hemisphere. These intersect along a band along the equatorial region
and their union gives the sphere surface as an open set.  

A standard method of studying the topology of a space $X$ is to associate sets of groups $G_i(X)$ to the space $X$ in such a  way that if $G_i(X)\neq G_i(Y)$ for a single group of the collection then the spaces $X,Y$ are not homeomorphic. Here two groups are equal if they are isomorphic. This is the subject matter of algebraic topology. There are many ways of associating groups to spaces. $K$-groups are one example that are useful for studying the topology of spaces called fiber bundles\cite{NashSen}. 

In physics, topological features often show by when boundary conditions are imposed. Thus if we consider a disc and consider a function that vanishes on its boundary it effectively means that all boundary points of the disc can be identified. As a result the disc is effectively equivalent to sphere.
Thus the function on the disc inherits topological properties associated with the surface of a sphere, such as the notion of winding maps. If the function of interest is valued in $\SU(2)$
then from the topological perspective we have a map $\SU(2)\longrightarrow \sph^2$. 
But an  $\SU(2)$ matrix, generically is of the form
$\left(\begin{smallmatrix}
a & b\\ -b^* & a^*
\end{smallmatrix}\right)$,
and has unit determinant, that is, $aa^{*}+bb^{*}=1$, where $(a^{*},b^{*})$
are the complex conjugates of $(a,b)$. Thus, the
 parameter values  of an $\SU(2)$ matrix range over $\sph^3$.
Hence in this case there is hidden topological feature associated with the map 
$\sph^3\longrightarrow \sph^2$. This map has been studied in topology and is known as the Hopf map. It is non trivial and leads to observational consequences such as the anomalous Hall effect \cite{bruno}. 

It is possible to prove that the different maps $\phi_i:\sph^n\rightarrow \sph^m$ between sphere surfaces of different dimensions can be combined to  form a group known as homotopy groups, $\pi_{n}(\sph^m)$. The combination of these maps makes use of the idea of homotopy equivalence and they always require one starting fixed point for the maps, called its base point\cite{NashSen}. In general  maps from a spaces $\sph^n$ to an arbitrary space $X$ give rise to homotopy groups $\pi_n(X)$ that characterize the topological properties of $X$. These groups have been widely studied. A topologically trivial space $X$ will have homotopy groups that are identity group for all $\sph^n$ maps, i.e. $\pi_n(X)=0$ for all $n$ values. 

There are many other groups that are used to characterize the topological properties of arbitrary spaces $X$ such as homology groups and cohomology groups. We will mention them in passing without going into the details of their definitions. Intuitively the homology groups study topological properties of a space with the help of boundary operators. The idea is that a boundary operator can identify the boundary of a region. If we find a boundary but there is no associated space we will have spotted a hole of the space which is a topological property of a space. 

One way of carrying out these calculations is by describing a given space in terms of elementary building blocks, such as points, lines, triangles. The boundary operator then maps 
each of these units to its boundary. Groups structures are introduced by maps between the geometric building blocks and abelian groups.

Cohomology groups are dual to homology groups, which means an element of a cohomology group can be paired with an element of a homology group to give a scalar. The dimension of a homology group is a number called a Betti number. These groups and ideas are discussed in, for example, \cite{NashSen,egh}. Our main focus will be on defining and calculating 
$K$-groups.

We  have seen that imposing vanishing boundary conditions on functions converts  a space 
$\R^n$, Euclidean $n$-dimensional spaces to the space $\sph^n$, an  
$n$-dimensional spheres with topological properties. Since functions in physics  very often are assumed to vanish at boundary points the main topological focus in  physics is to consider topological features that follow for functions defined on different dimensional spheres.  

The study of the topological properties of a physical system involve two spaces. The first is the sphere surface and the second is the physical property of interest. If for instance we are interested in the the flow of air on the surface of the earth the flows at a given point can be described by a velocity function, a vector function in three dimensions. Thus the two spaces required to describe air flows on the earth's surface are, a sphere surface, representing the earth and three dimensional Euclidean space glued at each point on the sphere to describe the air flow.
What we have described is a fiber bundle space. In such a space two spaces are glued together and a set of rules for gluing are introduced so that the system has continuity properties. We will return to properly define a fiber bundle later.

However this leaves out another widely studied class of spaces of physics. These are spaces that have periodicity. For a rectangle imposing periodicity in the two directions produces a topological torus. Thus tori of different dimensions are a natural class of spaces of interest in physics. Now we have a fiber bundle where one of the spaces is a torus. However to study such bundles require additional mathematical tools. The appropriate group for the problem is the $K$-group. These are the groups we study presently. We will study the special example of the topological insulator to illustrate the approach. 

 Topological insulators are condensed matter systems with periodic structures. Thus tori appear as the base space. Second the system has electrons. These are represented as spinors wavefunctions of the fiber space. The spin of the electrons, located at lattice sites, are in molecules and they interact strongly with orbital angular momenta present in the orbitals of the system and have discrete time-reversal symmetry. We will explain these terms and then show how a system with these properties can be a bulk insulator that have conducting surface points because of the topological properties of the system. The intuitive idea is that time-reversal invariance introduces topological ``twists" on the surface that lead to zero energy excitations. An insulator is a system where conduction of charges cannot happen because the ground state electron energy states, that are bands, due to the periodicity of the system, are all occupied and there is a energy gap between this ground state and the conduction band. Thus at temperature values that are not high enough to mix the ground state to the conduction band there cannot be conduction of charges. 

We show that when the $K$-groups is non zero the energy gaps vanish, for only special surface surface points, but not for bulk bands  due to topology, and conduction is possible, at these special points. In our demonstration we will show how to model a many body system with  strong spin-orbit interactions that has time-reversal invariance and show it can be done by considering the topological properties of a single relativistic Dirac equation of the surface of a torus. This is an unexpected result as the starting many body system considered is non relativistic.


We  start by summarizing features of many body physics we need in 
section~\ref{many-particle} and explain consequences of time-reversal 
invariant crystalline systems in section~\ref{sec:topology} to show how 
the many body problem can be reduced to the simpler problem of studying the 
topological properties of a Dirac particle on the surface of a torus. 
In sections \ref{sec:basic} and \ref{sec:kth} 
the basic relevant topological ideas, including those of $K$-theory, 
are discussed. 
In section \ref{sec:comput} we review the general method of calculating 
$K$-groups for bundles on tori and outline applications that include the topological insulator
in section.The existence of gap-less points as consequences of topology
 was realized by a number of authors \cite{Mele1,Mele2,fu} 
and the possibility of such systems to exist was anticipated 
theoretically \cite{Zhang1,Zhang2}. 
Observation of such a material with novel
physical properties \cite{Hassan} has led to the expectation 
that it is possible to theoretically identify  
 materials with unusual properties using 
topological arguments \cite{Zhang3}.
We compute the $K$-groups of time-reversal invariant systems in 
section \ref{sec:topinsul} before concluding in
section \ref{sec:concl} with a few remarks about the approach.
\section{Many particle systems}
\label{many-particle}
Let us start at the physics end by
explaining how quantum field theory enables us
model the electronic properties
of many body systems by a single Dirac equation.

\subsection*{Quantum Field Theory}
\label{QFT}
Quantum field methods are used in condensed matter physics where atoms and molecules are close together and quantum ideas become relevant. Here classical fields are replaced by quantum operators and individual classical particles are replaced by operator fields. Our aim is to show why many body condensed matter systems require quantum fields for their description and then show why the properties of such an interacting many system with strong spin-orbit coupling 
can be modeled by massive Dirac particles in a gauge field. Our starting system is non relativistic but the topological properties of system can be effectively analyzed by the relativistic Dirac equation. 

Let us consider a collection of $N$ identical  non-interacting, non-relativistic particles
each of mass $m$ and momenta ${p}_i$.
There are two very different ways of determining the total energy of such a 
quantum system. 
The first way is to find the energy eigenvalue  $E_N$ of the 
$N$-particle system by solving the Schr{\"o}dinger's equation,
\begin{equation}
-\frac{\hslash^2}{2m} \sum_{i=1}^{N}\nabla_i^2 \Psi(\x_1,\x_2,\cdots,\x_N)=E_N
\Psi(\x_1,\x_2,\cdots,\x_N),
\end{equation} 
where $\x_i\in\R^D$ denotes the position of the $i$-th particle in the
$D$-dimensional Euclidean space and $\nabla^2_i$ denotes the Laplacian in
$\x_i$.
The eigenvectors, called wave functions, are square-integrable complex
functions in $L^2(\R^D,\C)$.
The second way involves two steps. In the first step the Schr{\"o}dinger equation
is solved for every single particle \ie 
\begin{equation}
-\frac{\hslash^2}{2m} \nabla^2_i \psi(\x_i)=E_i \psi(\x_i),
\end{equation}
where the wave functions now correspond to single particles, in 
$\otimes_1^N L^2(\R,\C)$.
In the second step the fact that the $N$ particles are  
identical and non-interacting
is used by observing that each particle will have as its energy one of the 
allowed single particle energy eigenvalues of $E_i$.  
Since the particles are non interacting the total energy is  simply
\begin{equation}
\label{nE}
E_N=\sum_{i=1}^{N}n_i E_i,
\end{equation}
where $n_i$ is the number of particles among the $N$ that have energy $ E_i$,
that is  $\{n_i|\sum_{i=1}^N n_i=N\}$  is a partition of $N$ of length $N$.
There seems, at this stage, very little to recommend the
second method. It needs us to anyway solve Schr{\"o}dinger's equation  
and it seems to be have rather limited scope as the total energy of a 
system of $N$ objects is  equal to the sum of the energies of the 
objects separately only when there are no interactions. 

But interactions are the heart of physics.  
Interactions may be incorporated in the first approach rather easily by
adding a term corresponding to the potential energy of the particles due to
the interactions, thereby modifying the equation to 
\begin{equation}
\left[-\sum_{i=1}^{N}\frac{\hslash^2}{2m} \nabla_i^2+  
\sum_{i=1}^N \sum_{j=1}^N V(\x_i, \x_j)\right] 
\Psi(\x_1,\x_2,\cdots,\x_N)=E_N\Psi(\x_1,\x_2,\cdots,\x_N),
\end{equation}
solving which will give us $E_N$. 

Including interactions in the second approach, however, seems to be 
hopeless as the sum of terms cannot be extended to include interactions
while retaining the product structure of  
the eigenvectors in $\otimes_1^NL^2(\R,\C)$.  
This  problem  can be solved by reinterpreting \eq{nE} as we now show by using 
ideas of quantum field theory.
We start by considering the elements of the quantum oscillator algebra, defined in terms
of generators that satisfy 
\begin{equation}
\label{oscl:alg}
\begin{split}
[a_i, a_j^{\dagger}]=\delta_{i, j},\quad
[a_i, a_j] = [a_i^{\dagger}, a_j^{\dagger}] = 0,
\end{split}
\end{equation} 
where $[A, B]=AB-BA$ is the commutator and $AB$ denotes the composition of
$A$ with $B$. This algebra makes it appearance, for example, in the algebra
of raising and lowering operators for the harmonic oscillator in quantum
mechanics. But here they are interpreted in a different way. Let us consider the operators 
\begin{equation}
\label{numop}
\N=\sum_{i=1}^{N}\hat{N}_i,\quad
\hat{N}_i = a_i^{\dagger}a_i.
\end{equation} 
Using this definition and \eq{oscl:alg} as well as the identity,
$[AB,C]=A[B,C]+[A,C]B$, for any three operators $A$, $B$, $C$, 
we have 
\begin{lemma}
\label{lem1}
\begin{equation}
[\N_i, a_j] = -a_j\delta_{ij},\quad
[\N_i, a_j^{\dagger}] = a_j^{\dagger}\delta_{ij}.
\end{equation}
\end{lemma}
Let $|n\rangle$ be an eigenvector of $\N_i$ with eigenvalue $n_i$, that is, 
\begin{equation}
\label{e:Ni}
\N_i|n\rangle=n_i|n\rangle.
\end{equation}
Defining a state 
$|s\rangle = a_i|n\rangle$, we have 
\begin{equation}
\langle s|s\rangle
=\langle n|a_i^{\dagger}a_i|n\rangle
=\langle n|\N_i|n\rangle
=n_i\langle n|n\rangle.
\end{equation} 
Semi-positivity of inner products of states in a Hilbert space then leads to
\begin{lemma}
\label{lem2}
Eigenvalues of the operator $\N_i$ and hence of $\N$ are non-negative. 
\end{lemma}
For an eigenvector $\ket{n}$ of $\N_i$ as in \eq{e:Ni}, we have, using
Lemma~\ref{lem1}, 
\begin{equation}
\begin{split}
-a_j\delta_{ij}|n\rangle
&=[\N_i,a_j] |n\rangle \\
&=(\N_i a_j - a_j\N_i)|n\rangle \\
&=\N_i a_j|n\rangle - a_jn_i\delta_{ij}.
\end{split}
\end{equation} 
This and a similar manipulation with $a^{\dagger}$ leads to 
\begin{lemma}
\label{lem3}
If $|n\rangle$ is an eigenvector of $\N_i$ with eigenvalue $n_i$, then
$a_i|n\rangle$ and $a_i^{\dagger}|n\rangle$ are eigenvectors of $\N_i$ with
eigenvalues $n_i-1$ and $n_i+1$, respectively, that is,
\begin{gather}
\N_ia_j|n\rangle = (n_i-1)a_j\delta_{ij}|n\rangle,\\
\N_ia_j^{\dagger}|n\rangle = (n_i+1)a_j\delta_{ij}|n\rangle.
\end{gather} 
\end{lemma}
In the light of these $a_i$ and $a_i^{\dagger}$ are called \emph{lowering}
and \emph{raising} operators, respectively. Together they are also referred to
as \emph{ladder} operators. The lowering of the eigenvalue of $\N_i$ by $a_i$
Reduction of the eigenvalue of $\N_i$ by $a_i$ can not be allowed
indefinitely as this will eventually contradict Lemma~\ref{lem2}. In order to
make the Lemma~\ref{lem2} and Lemma~\ref{lem3} compatible we assume the
existence of an eigenvector of $\N_i$ with a predefined non-negative
minimum eigenvalue, customarily taken to be zero. This is effected by the 
\begin{hypo}[Existence of vacuum]
In the Hilbert space of states of the many-particle system there exists a
vector, denoted $|0\rangle$, with unit norm,
such that $a_i|0\rangle=0$, for each $i$. This vector is referred to as the
\emph{vacuum}.
\end{hypo}
Clearly, $\N_i$ has a lowest eigenvalue, $\N_i|0\rangle=0$. 
This renders $\N_i$ the interpretation of a \emph{number operator}, with
non-negative numbers as eigenvalues. The vectors in the Hilbert space are
then many-particle states with the number of particles of type $i$ as 
$n_i$, often referred to as the
\emph{occupation number}. The raising and lowering operators are now regarded as \emph{creation} and \emph{destruction} operators for particles from the
vacuum. They change the number of particles in the vacuum state. Thus a simple many body system of free particles can be described by using quantum operators that carry a discrete index associated with the particle that can label its momentum, a continuous variable. Thus
this simple many body system can be described by operators that carry a continuous label: they are quantum fields. We thus have a state vector  for an assembly of particles states with labels $i$ given by,
\begin{equation}
|n\rangle = |n_1,n_2,\cdots, n_i,\cdots\rangle.
\end{equation} 
This is a state with occupation number $n_i$ for the $i$-th particle. We
assume it to be normalized to unity,
\begin{equation}
\langle n|n\rangle = 1.
\end{equation} 
Since
by Lemma~\ref{lem3} the destruction operator $a_i$ maps the state 
$|n_1,n_2,\cdots, n_i,\cdots\rangle$ to the state 
$|n_1,n_2,\cdots, n_i-1,\cdots\rangle$, writing 
\begin{equation}
a_i |n_1,n_2,\cdots, n_i,\cdots\rangle 
= \alpha_i |n_1,n_2,\cdots, n_i-1,\cdots\rangle, 
\end{equation} 
and considering the inner product of the state $a_i|n\rangle$ with itself,
we obtain
\begin{equation}
|\alpha_i|^2=\langle n|\N_i|n\rangle
=n_i\langle n|n\rangle = n_i.
\end{equation} 
The creation operators can be treated similarly. We thus have the vectors in
the Hilbert space normalized as
\begin{gather}
a_i |n_1,n_2,\cdots, n_i,\cdots\rangle 
= \sqrt{n_i} |n_1,n_2,\cdots, n_i-1,\cdots\rangle ,\\
a_i^{\dagger} |n_1,n_2,\cdots, n_i,\cdots\rangle 
= \sqrt{n_i+1} |n_1,n_2,\cdots, n_i+1,\cdots\rangle . 
\end{gather} 

To summarize, we have taken two important steps.
The first step was to show that for $N$ non-interacting particles
the total energy of the system can be written in terms of the eigenvalues 
of a collection of quantum field theory number operators which are built 
out of creation-destruction operators. 
The second step was to determine the way these  
creation destruction operators act on the number labeled eigenvectors 
introduced. 

We now take the third step which is to  show how the quantum field 
creation destruction operator approach is equivalent to the $N$ free 
particle Schr{\"o}dinger approach. This establishes why quantum field is relevant in condensed matter physics.  We state the correspondence in the 
form of two Equivalence Theorems
\begin{theorem}
Let $\hat{H} =\sum_{i=1}^{N} E_i a_i^{\dagger} a_i$ and 
$\hat{N}=\sum_{i=1}^{N}a_i^{\dagger}a_i$ denote the Hamiltonian and the 
number operator, respectively,
with $E_i=\frac{p_i^2}{2m}$. Suppose,
\begin{gather}
\hat{H}|N,E_{N}\rangle=E_{N}|N, E_{N}\rangle,\\
\hat{N}|N, E_{N}\rangle=N|N, E_{N}\rangle,
\end{gather}
 and there is a state $|0, 0\rangle$ such that $\hat{H}|0, 0\rangle=\hat{N}|0, 0\rangle=0$ then the system, described  here in terms of momenta labels $i=p_i$ has an equivalent position space representation given by:
\begin{gather*}
\hat{H}=-\frac{\hslash^2}{2m} \int dx \Psi^{\dagger}(x) \nabla^2 \Psi(x),\\
\hat{N}= \int dx \Psi^{\dagger}(x) \Psi(x),
\end{gather*}
where $\Psi^{\dagger}(x), \Psi(y)$ are operators which obey the commutation relations
\begin{equation}
 [\Psi^{\dagger}(x), \Psi (y)]=\delta (x-y),
\end{equation}
\begin{equation}
 [\Psi^{\dagger}(x), \Psi^{\dagger}(y)]=[\Psi(x), \Psi(y)]=0,
\end{equation}
where  $\Psi(y)=\int dp e^{-ipy}a_p$ , and 
$\Psi^{\dagger}(x)=\int dp e^{ipx}a^{\dagger}_p$ i.e the position 
labeled operators  $\Psi^{\dagger}(x), \Psi(y)$ are simply the Fourier 
transforms of the momentum labeled operators $a^{\dagger}_p, a_p$. 
\end{theorem}
The commutation relations written down follow from those of the creation destruction operators introduced  earlier. We can intuitively think of $\Psi^{\dagger}(x)$ as creating a particle at the point $x$ and $\Psi(x)$
as destroying a particle at the point $x$.

 Let us show how $[\Psi^{\dagger}(x), \Psi(y)]=\delta(x-y)$ comes from $[a^{\dagger}_p, a_q]=\delta_{p,q}$. The other results can be proved in the same way. 
We use Fourier transforms to write
\begin{gather*}
\Psi(y)=\int dq e^{-iqy} a_q,\\
\Psi^{\dagger}(x)=\int dp e^{ipx} a^{\dagger}_p.
\end{gather*}
Then
\begin{equation}
[\Psi^{\dagger}(x), \Psi(y)]=\int dq \int dp e^{ipx-iqy}[a^{\dagger}_p,a_q].
\end{equation}
Using $[a^{\dagger}_p,a_q]=\delta_{p,q}$ we get
\begin{equation}
[\Psi^{\dagger}(x),\Psi(y)]=\int dp e^{ip(x-y)}=\delta(x-y),
\end{equation}
which established the result. 

Next, we link the operator approach to the Schr{\"o}dinger equation approach by
defining a function $\Phi(\x_1, \x_2,\cdots,\x_N)$ as follows.
\begin{theorem}[Equivalence Theorem]
\begin{equation}
\frac{1}{\sqrt{N!}}\langle 0,0|\Psi(\x_1) \Psi(\x_2)\cdots\Psi(\x_N)|E_N, N\rangle
=\Phi(\x_1, \x_2,\cdots,\x_N).
\end{equation}
We next show that $\Phi(\x_1,\x_2,\cdots,\x_N)$ satisfies Schr{\"o}dinger's equation. We thus have a bridge connecting the two approaches.
\begin{equation}
-\frac{\hslash^2}{2m} \sum_{i=1}^{N}
\nabla_i^2 \Phi(\x_1,\x_2,\cdots,\x_N)=E_N \Phi(\x_1,\x_2,\cdots,\x_N).
\end{equation}
\end{theorem}
\begin{proof}
The only tool at our disposal in the operator formulation are the commutation relations that are given. Thus in order to derive results from the operator approach we have to use the commutation relations given. We proceed to show 
how this is done.  First we establish a simple result.  We note that
\begin{equation}
\frac{1}{\sqrt{N!}}\langle 0, 0|\Psi(\x_1) \Psi(\x_2)\cdots\Psi(\x_N)
\hat{H}|E_N, N\rangle=E_N \Phi (\x_1,\x_2,\cdots,\x_N).
\end{equation}
This step gives us the right hand side of Schr{\"o}dinger's equation. To show that the left hand side has Schr{\"o}dinger's equation hidden in it  we must use the
commutation relations which define the operator approach. The commutation relations appear when include the information that $\hat{H}|0, 0\rangle=0$.  This can be added on to the equation written by simply replacing the product
$\langle 0, 0|\Psi(\x_1)\Psi(\x_2)\cdots\Psi(x_N) \hat{H}|E_N, N\rangle$ by the commutator as follows 
\begin{equation}
\langle 0, 0|\Psi(\x_1)\Psi(\x_2)\cdots\Psi(x_N) 
\hat{H}|E_N, N\rangle=\langle 0, 0|[\Psi(\x_1),\Psi(\x_2)\cdots\Psi(\x_N), 
\hat{H}]| E_N, N\rangle,
\end{equation}
because on the left hand side we have the vector $|0, 0\rangle$ and $\hat{H}|0, 0\rangle=0$  the term of the commutator with $\hat{H}$ on the left hand side gives zero contribution. 
\end{proof}
We next evaluate the commutator and show that we get Schr{\"o}dinger's equation.To do this we need to use the expression for $\hat{H}$ in terms of the operators $\Psi^{\dagger}(x), \Psi(x)$, and we need the 
\begin{lemma}[Calculating Lemma]
Given a collection of operators $A_1, A_2,\cdots,A_N, B$, the commutator
\begin{equation}
[A_1A_2\cdots A_N, B]=\sum_{i=1}^{N}A_1\cdots A_{i-1}[A_i, B] A_{i+1}\cdots A_N.
\end{equation}
\end{lemma}
This result can be proved by induction. First one checks that the result is true for $N=2$, then it is assumed to hold for $i-1$ and from this it is shown
that the result holds for $i$.  Let us proceed to apply this lemma to our problem. We have
\begin{equation}
[\Psi(\x_1) \Psi(\x_2)\cdots\Psi(\x_N), \hat{H}]
=\sum_{i=1}^{N} 
\Psi(\x_1)\cdots\Psi(\x_{i-1}[\Psi(\x_i), \hat{H}] 
\Psi(\x_{i+1}\cdots\Psi(\x_N).
\end{equation}
Thus we need to work out the commutator $[\Psi(\x_i), \hat{H}]$.
Using the lemma again we get
\begin{equation}
-\frac{\hslash^2}{2m} [\Psi(x_i), \int dy \nabla^2_y \Psi(y)]=-\frac{\hslash^2}{2m}\nabla^2_i \Psi(x_i).
\end{equation}
In order to get this result we used the fact that $[\Psi^{\dagger}(y), \Psi(x_i)]=\delta(y-x_i)$ and all other commutators  vanish. Using this result the left hand side of Schr{\"o}dinger's equation is obtained. Since
\begin{equation}
\sum_{i=1}^{N}\langle 0, 0|\Psi(x_1) \Psi(x_2)\cdots\Psi(x_{i-1}[\Psi(x_i), \hat{H}] \Psi(x_{i+1})\cdots\Psi(x_N)|E_N, N\rangle
\end{equation}
\begin{equation}
=-\sum_{i=1}^{N!}\frac{\hslash^2}{2m}\nabla^2_i\langle 0,0| \Psi(x_1) \Psi(x_2)
\cdots\Psi(x_i)\cdots\Psi(x_n)| E_N, N\rangle,
\end{equation}
where $\nabla^2_i$ means that only functions of the variable $x_i$ are acted
on by the differential operator.

Let us comment on what has been done. The connection bridge suggests a way
in which interactions can be included in the quantum field theory approach  as we now describe. Furthermore once interactions  are introduced properly they give can be used to construct $N$ particle Schr{\"o}dinger wavefunctions. The two approaches
are equivalent.   We established that the
non interacting term model represents non interacting part of Schr{\"o}dinger's equation. Let us now interpret this result as follows. The $\Psi^{\dagger}(x) \Psi(x)$  factor can be regarded as the ``particle density operator" $\rho(x)$. Then this term is simply the ``energy" $-\frac{\hslash^2}{2m} \nabla^2$  associated with a single particle.  This suggests that interaction energies should be associated with ``two- particle densities" that depend on two coordinates $x, y$ and the potential energy $V(|x-y|)$ associated of the two  particles. Thus a term of the form
\begin{equation}
\frac{1}{2}\int dx \int dy \Psi^{\dagger}(x) \Psi^{\dagger}(y) V(|x-y|) \Psi(x) \Psi(y)
\end{equation}
would seem plausible. Once this idea is grasped ``three- particle densities"
leading to three-body forces and so on can be considered. It is possible to
show that the expression for the potential energy contribution written down
does, in fact, lead to $\Phi(x_1, x_2,\cdots,x_N)$ 
satisfying Schr{\"o}dinger's equation
with the potential $V(|x-y|)$.  We will not establishing this technical result
but this discussion should make it clear that quantum field theory does provide
a natural framework for condensed matter physics.  The problem is to
work out physical properties of a system within this framework.
\subsubsection*{Fermions}
The presentation of quantum field theory given  so far has been for bosons, i.e particles which have 
multi-particle wave-functions symmetric under the exchange of particles. This
is true for the wave-function 
\begin{equation}
\Phi(x_1, x_2,\cdots,x_N) =\frac{1}{\sqrt{N!}}\langle 0, 0|\Psi(x_1)\Psi(x_2)
\cdots\Psi(x_{i})\Psi(x_{i+1})\cdots\Psi(x_N)|E_n, N\rangle.
\end{equation}
Since the operators $\Psi(x_{i})$ and 
$\Psi(x_{i+1})$ commute. To describe fermions
we need to have multi-particle wave functions that are antisymmetric under 
particle the exchange of particles. All half integer normal spin systems, such as electrons, are fermions while integer spin systems are bosons.  A natural way of constructing antisymmetric multiple particle wave functions is to introduce quantum field operators that anticommute. Taking the corresponding $N$-particle wavefunction to be of the same form as $\Phi(x_1, x_2,\cdots,x_N)$ but with operators $\Psi(x_{i})$ that now anticommute i.e satisfy the anticommutation rules,
\begin{gather}
[\Psi^{\dagger}(x), \Psi(y)]_{+}=\delta(x-y),
\\
[\Psi^{\dagger}(x), \Psi^{\dagger}(y)]_{+}=[\Psi(x), \Psi(y)]_{+}=0,
\end{gather}  
where $[A, B]_{+}=AB+BA$ is called the anticommutator,  makes  
$\Phi(x_1, x_2,\cdots,x_N)$ antisymmetric under particle exchange.  It
is possible to show, following the steps described for bosons, that 
$\Phi(x_1,x_2,\cdots,x_N)$ satisfies the $N$-particle Schr{\"o}dinger's 
equation. However in order to evaluate $[\hat{H}, \Psi(x_i)]$ the identity 
\begin{equation}
[AB, C]=A[B,C]_{+}-[A,C]_{+}B
\end{equation}
relating commutators to anticommutators, 
valid for any three operators A, B and C, has been used.
The momentum space anticommutation relations for fermions which follow
from the position space results written down are
\begin{gather}
[a^{\dagger}_p, a_q]_{+}=\delta_{p, q},a\\
[a^{\dagger}_p, a^{\dagger}_q]_{+}=[a_p, a_q]_{+}=0.
\end{gather}
An immediate consequence of these anti commutation relations is the important result for the allowed eigenvalues of the number operator $n_p=a^{\dagger}_pa_p$
\begin{equation}
(n_p)^2=a^{\dagger}_pa_pa^{\dagger}_pa_p=a^{\dagger}_p(1-a^{\dagger}_pa_p)a_p=a^{\dagger}_pa_p=n_p.
\end{equation}
From this result $(n_p)^2=n_p$ it follows that the eigenvalues of $n_p$ can only be either zero or one. Thus only one fermion can occupy a given momentum state.

Let us next describe a system with strong spin-orbit forces and time-reversal invariance
in mathematical terms and show that such a system can be described by the relativistic Dirac equation in a gauge field. Recall the electromagnetic field is a gauge field called the electromagnetic vector potential. Mathematically a system with spin is treated as
a two component spinor in non-relativistic physics. Thus in the N-body interacting
system with spin-orbit interaction will be described by a $N$-body Hamiltonian and an associated N-body Schr{\"o}dinger equation with the spin-orbit interaction modeled as 
$\sum g\sigma_iL_i$ with $\sigma_i$ representing the spin of the $i^{th}$ particle and $L_i$ representing the orbital moment of the particle. The factors $\sigma_i$ are two-by-two matrices known as Pauli spin matrices 
\cite{NashSen}.Thus  in such a formulation the spin-orbit interaction 
term is written as 
$\sum \sigma_i L_i a^{\dagger}_{k,i,\sigma_i}a_{k,i,\sigma_i}$, 
the kinetic energy term as  $\sum \frac{k^2_i}{2m_i} 
a^{\dagger}_{k,i,\sigma_i}a_{k,i,\sigma_i}$,
while the interaction term has the structure $\sum a^{\dagger}_{k_1,
i_1,\sigma_{i_1}}a^{\dagger}_{k_2, i_2, \sigma_{j_2}}V(k_1,k_2,k_3,k_4, 
\sigma_{i_1},\sigma_{i_2},\sigma_{i_3},\sigma_{i_4})a_{k_3,\sigma_{i_3}}
a_{k_4,\sigma_{i_4}}$. 
In intuitive physical terms these different terms are easily understood. For instance
the combination $a^{\dagger}a$ represents the number of particles
with specific values for the momenta and spin. Thus the spin-orbit interaction energy term, in this formalism,  is the simple statement that $a^{\dagger}a$ particles have  energy $\sigma.L$ the Hamiltonian.  Here we have written terms of the Hamiltonian using  momentum space i.e in terms of the reciprocal lattice variables expressed as momenta known as the Brillouin zone. The Brillouin zone represents the basic region of momentum space defined by the periodicity properties of the system. It is topologically a torus.

We next make a general remark about systems with strong spin-orbit interaction.
In the low momentum region the dominant terms will be the spin-orbit and interaction
potential terms. Neglecting the interaction potential terms, for the moment, we see our system then becomes a sum of $N$ independent spin-orbit interaction terms. Now $L_i$ is a representation of $\SO(3)$, which, in terms of spinors, it is a representation of $\frac{\SU(2)}{\Z_2}$. This orbital angular momentum term multiplied by $g(x)$ thus has the structure of a gauge field. The label $x$ indicates the site of the molecule in which the electron is located. A natural step is to rewrite this gauge field in a ``gauge'' invariant way as 
$\sigma\cdot(k-gL)$, where $k$ is lattice momentum. But this is exactly the form of a mass-less Dirac equation.  The effect of the potential terms, neglected so far, in the low momentum region, can be included by simply making the Dirac equation massive. There are standard methods of quantum field theory that can be used to establish this result.  Thus simple but general argument suggests that a system with strong spin-orbit interactions , in the low momentum region, can be effectively represented by a collection of massive Dirac equations with its momentum space limited to a torus, called a Brillouin zone. In this step the spatial lattice periodicity is replaced by a corresponding induced  lattice periodicity in momentum space called the reciprocal lattice. Thus  non-interacting particles massive Dirac particles with the angular momentum, interpreted as a gauge field, provide a good description of the system for low momenta values. They encode lattice periodicity and the presence of strong spin-orbit interactions. 

We have at this stage introduced the physics concepts required to discuss the topological insulator. It is a system with electron spins in a periodic lattice. The electrons experience strong spin-orbit interactions and the system has time-reversal invariance. The periodicity gives rise to energy band structures that do not allow conduction in the bulk but there are special surface points in the system where there is conduction. The mathematical problem is to show that the physical properties of a topological insulator follow from the topological properties of a Dirac equation in a gauge field in a lattice system which is the three torus $\sph^1\times \sph^1\times \sph^1$ in the bulk and a two torus  $\sph^1\times \sph^1$ on its surface
and the fact that the system has time-reversal invariance. We established these results using the $K$-groups. The key idea is a non vanishing $K$-group implies the vanishing of the gap between a non-conducting and a conducting band. All our discussions will be in the momentum space lattice induced from the spatial lattice of the system. 
\section{Time reversal symmetry in crystals}
\label{sec:topology}
We have considered the effect of strong spin-orbit interactions. Now we turn to study the topological consequences of time-reversal symmetry. For this we examine the general case of electronic states of a crystalline system. The eigenfunctions of a single particle system is described by the elliptic self-adjoint Hamiltonian in a periodic lattice,
\begin{equation}
H = -\nabla^2 + V(x),
\end{equation} 
where $V$ is a sufficiently regular real-valued periodic function on $\R^D/\G$. As $H$
is invariant under $\trgr$, it possesses the same eigenfunctions as the
translations in $\trgr$. 
Introducing  an orthogonal set of basis vectors in the Hilbert space
$\mathcal{H}$ underlying the irreducible representations of $\trgr$ 
\begin{gather}
\Psi_{n}(x,k)=e^{i\langle  k,\gamma\rangle}\psi_{n}(x,k),\\
\psi_{n}(x+\gamma,k)=\psi_{n}(x,k), \quad\gamma\in\G,
\end{gather}
called Bloch functions, the electronic states are given by a solution to the
eigenvalue problem 
\begin{equation}
H(k)\psi_n(x,k) = E_n(k)\psi_n(x,k).
\end{equation}  
The eigenvalue $E_n(k)$ of the Bloch Hamiltonian $H(k)$, 
called the $n$-th band function, is a continuous
function of the crystal momentum $k\in\B$ for each $n=0,1,2,\cdots$.
 
Topological insulators are crystalline systems invariant under
time-reversal. Electronic states of such systems, in addition to being
irreducible representations of $\trgr$, form a representation of the
finite group generated by the time-reversal operator $\T$. 
Assuming that the time evolution of an eigenstate $\psi_n(x,k)$ 
of the Hamiltonian $H$ with eigenvalue $E$ is given by 
$e^{iEt}\psi_n(x,k)$, with $E$ positive, in conformity with the
time-dependent Schr\"odinger equation, the 
time-reversal operator is taken to be 
anti-unitary, that is an endomorphism of the Hilbert space $\mathcal{H}$
mapping the inner product to its complex conjugate. 
It changes the sign of the momentum and time, keeping the position of a state
unaltered, that is
\begin{equation}
\T: \psi_n(x,k)\map\psi^{\star}_n(x,-k), 
\end{equation} 
where a $\star$ denotes complex conjugation. On physical grounds 
a state is assumed to come back to itself up to a phase if
acted on by $\T$ twice, so that
\begin{equation}
\label{Tsq}
\T^2 = \alpha, \quad |\alpha|=1.
\end{equation} 
The time-reversal operator $\T$ is assumed to act on the Bloch Hamiltonian
by conjugation as
\begin{equation}
\label{T:H}
\T H(k)\T^{-1}=H(-k), 
\end{equation} 
The Hilbert space $\mathcal{H}$ then correspond to a chosen representation 
of $\T$. 

If the  spin of electrons is taken into
consideration, then the Hamiltonian contains terms involving spin $1/2$
operators valued in a representation of the group $\SU(2)$ furnished by 
the $2\times 2$ Pauli matrices.
The corresponding Bloch wavefunctions over $\B$, written as 
$\psi_n(x,k) = \left(\begin{smallmatrix}\psi_{n\uparrow}(x,k)\\
\psi_{n\downarrow}(x,k)\end{smallmatrix}\right)$, transform as vectors
under the $\SU(2)$,
where the upper and lower components are referred to as the
spin up and down states, respectively. 
By flipping the sign of momentum $\T$ changes the sign of 
angular momentum. The same is assumed for spin.

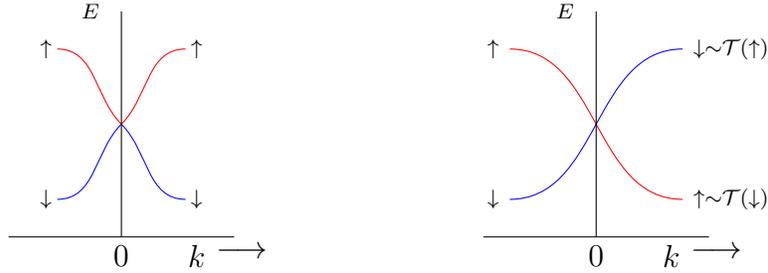
\begin{figure}[h]
\begin{center}
\begin{tikzpicture}
\begin{scope}
\node (X) at (0,2) {}; 
\node[label=west:{$\scriptstyle E$}] (Y) at (0,3.5) {}; 
\node[label=center:{$\scriptstyle\downarrow$}] (A) at (1,1) {}; 
\node[label=center:{$\scriptstyle\uparrow$}] (B) at (1,3) {}; 
\node[label=center:{$\scriptstyle\downarrow$}] (C) at (-1,1) {}; 
\node[label=center:{$\scriptstyle\uparrow$}](D) at (-1,3) {}; 
\draw [blue] (A.west) to [out=180,in=315] (X.center); 
\draw [red] (B.west) to [out=180,in=45] (X.center); 
\draw [blue] (C.east) to [out=0,in=225] (X.center); 
\draw [red] (D.east) to [out=0,in=135] (X.center); 
\draw ($(X)-(0,1.5)$) to (Y.center);
\draw ($(X)+(-1.5,-1.5)$) to ($(X)+(1.5,-1.5)$);
\node (k0) at ($(X)-(0,1.75)$) {$ 0$};
\node (k) at ($(X)+(1,-1.75)$) {$k$};
\node (arr) at ($(X)+(1.6,-1.7)$) {$\rt$};
\end{scope}
\end{tikzpicture}
\hskip 1in
\begin{tikzpicture}
\begin{scope}
\node (X) at (0,2) {}; 
\node[label=west:{$\scriptstyle E$}] (Y) at (0,3.5) {}; 
\node[label=east:{$\scriptstyle\uparrow\sim\T(\downarrow)$}] (A) at (1,1) {}; 
\node[label=east:{$\scriptstyle\downarrow\sim\T(\uparrow)$}] (B) at (1,3) {}; 
\node[label=west:{$\scriptstyle\downarrow$}] (C) at (-1,1) {}; 
\node[label=west:{$\scriptstyle\uparrow$}](D) at (-1,3) {}; 
\draw[red] (D.west) to [out=0,in=180] (A.east);
\draw[blue] (C.west) to [out=0,in=180] (B.east);
\draw ($(X)-(0,1.5)$) to (Y.center);
\draw ($(X)+(-1.5,-1.5)$) to ($(X)+(1.5,-1.5)$);
\node (k0) at ($(X)-(0,1.75)$) {$ 0$};
\node (k) at ($(X)+(1,-1.75)$) {$k$};
\node (arr) at ($(X)+(1.6,-1.7)$) {$\rt$};
\end{scope}
\end{tikzpicture}
\end{center}
\caption{Osculatory (left) and intersecting (right)
band functions for Kramer pairs}
\label{fig:Cross}
\end{figure}
A consistent choice of $\T$ is given by the transformation 
\begin{equation}
\label{Kpair}
\T :\ \begin{pmatrix} 
\psi_{n\uparrow}(x,k)\\\psi_{n\downarrow}(x,k)\end{pmatrix}\map  
\begin{pmatrix} \psi^{\star}_{n\downarrow}(x,-k)\\
-\psi^{\star}_{n\uparrow}(x,-k)\end{pmatrix}.
\end{equation} 
with $\T^2=-1$. 
The identification of wavefunctions up to a sign under $\T^2$ 
breaks the group $\SU(2)$ acting on them to $\SU(2)/\Z_2\simeq \SO(3)$.
Invariance under time-reversal requires the existence of a state with
energy $E(-k)$ if one with $E(k)$ exists, implying thereby, that
states with $E(k_0)=E(-k_0)$ at a point $k_0$, called the Kramer point.
At these points there are two  degenerate states,  $\psi$ and 
$\T\psi$, called a Kramer pair. 
Generically, $k_0=0$ is a Kramer point as well as $k_0=\pm 1$, as these two
points are identified in $\B$. 
A Kramer pair at $k=0$ may be associated to states on two sides
of $k=0$ in two inequivalent ways maintaining the continuity of Bloch
wavefunctions as well as the band function. This is schematically
shown in Figure~\ref{fig:Cross}, where band functions 
are plotted and the spinorial components of wavefunctions are indicated. 

In one configuration, shown in the diagram on the left, the spin up and 
down components of the Kramer pair go over respectively to the up and 
down components of a Bloch wavefunction on both sides, resulting into two
osculatory curves at the Kramer point. 
The second configuration is obtained as the up and
down components go over respectively to the up and down components on 
one side but to a time-reversed wavefunction on the other, with up and down
components exchanged according to \eq{Kpair}, as indicated in the diagram on
the right of Figure~\ref{fig:Cross}. This results in two intersecting curves
with distinct tangents at the Kramer point. Existence of distinct
tangents mean that the band function forms a Dirac cone at this point, since
the dispersion in a neighborhood of the intersection point can be brought to
the form $E=\pm k$. Hence even though the original system is non relativistic the
presence of time-reversal invariance and lattice periodicity lead to a Dirac equation structure when spin effects are important in the system.

The two types of band functions described are the only ones possible for a time
reversal invariant crystalline system.
\section{Wavefunctions and bundles}
\label{sec:basic}
In this section we describe Bloch wavefunctions on the Brillouin zone 
$\B$ in terms of fiber bundles, that have a torus as base space and a fiber on
which a group $G$ acts. These are $G$-bundles. 
A $G$-bundle \cite{NashSen,egh} over $M$ is
built by gluing together a fixed vector space at each point of 
a smooth manifold  $M$, called a \emph{base space}.
Each point of the vector space, called a \emph{fiber}, is acted upon 
linearly by a group $G$. The action of $G$ is  assumed to be transitive, 
meaning every point of a fiber $F$ can be reached from some point in 
$F$ through an action of $G$ and there is no fixed points under 
 the action of $G$. 
The gluing procedure is  described by 
covering the base space $M$ with a collection of possibly overlapping
contractible open sets $\{U_\alpha, U_\beta,\cdots\}$ with a choice of 
vector spaces  $\{F_{\alpha}, F_{\beta},\cdots\}$ on each open
set, where $\alpha, \beta,\cdots$ are valued in some countable index set. 
On an open set $U_\alpha$ the
bundle $V$ is simply the product of the spaces $U_\alpha$ and $F_{\alpha}$.
Thus at this stage we have a collection of local bits of the overall bundle space.
Each bit is trivial in the topological sense. Topology of the bundle comes from
the way these trivial bits are glued together. There is a projection map that maps 
the fiber space to a point on the base space and there is an inverse map from the base space 
to a bit of the bundle space. The gluing procedure is the key topological step. It glues together points that belong to two different open sets of $M$. They are called transition
functions. 
A $G$-bundle is said to be of \emph{rank} $k$ if the dimension of the fiber is  
$k$ at every point of $M$. Thus the bundle is constructed by gluing
together spaces, that are locally the product of 
one of the open sets that cover the base space $M$ and 
the fiber space $F$. The twists and turns of the resultant bundle space
are captured by the gluing procedure.
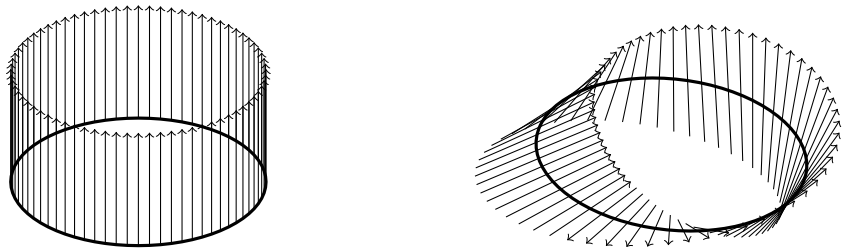
\begin{figure}[h]
\begin{center}
\tikzset{
	CylinderPersp/.style={scale=1.5,x={(-.8cm,-0.4cm)},y={(0.8cm,-0.4cm)},
    z={(0cm,1cm)}},MyPoints/.style={fill=white,draw=black,thick}}
\begin{tikzpicture}[CylinderPersp,font=\large]
	\def\h{.5}

	\foreach \t in {5,10,...,360}
		\draw[->] ({cos(\t)},{sin(\t)},0)
      --({cos(\t)},{sin(\t)},{2.0*\h});
	\draw[very thick] (1,0,0) 
		\foreach \t in {5,10,...,360}
			{--({cos(\t)},{sin(\t)},0)}--cycle ;
\end{tikzpicture}
\tikzset{
	MobiusPersp/.style={scale=1.8,x={(-.6cm,-0.4cm)},y={(0.8cm,-0.4cm)},
    z={(0cm,1cm)}},MyPoints/.style={fill=white,draw=black,thick}}
\hskip 1in
\begin{tikzpicture}[MobiusPersp,font=\large]
\foreach \t in {6,12,...,360} 
\draw [->]
({(1-0.5*cos(\t/2))*cos(\t)},{(1-0.5*cos(\t/2))*sin(\t)},{-0.5*sin(\t/2)}) --
({(1+0.5*cos(\t/2))*cos(\t)},{(1+0.5*cos(\t/2))*sin(\t)},{0.5*sin(\t/2)})  ;
	\draw[very thick] (1,0,0) 
		\foreach \t in {5,10,...,360}
			{--({cos(\t)},{sin(\t)},0)}--cycle ;
\end{tikzpicture}
\end{center}
\caption{Fiber Bundles: trivial (left) and non-trivial (right)}
\label{fig:fiber}
\end{figure}
The local product structure of a bundle $V$ is called a local trivialization. 
Thus, a $G$-bundle of rank $k$ over a smooth manifold $M$ is a
smoothly varying locally trivial family of $k$-dimensional vector spaces 
endowed with an action of $G$. If the bundle
$V$ can be described as the direct product of $M$ and a vector space
$F$ then the bundle is said to be \emph{trivial}.
As an example, if $M$ is a
circle $\sph^1$ and the fiber at each point is an oriented line, then the
lines can be glued together maintaining their orientations to form
a cylinder, which is a trivial bundle, or with a twist, 
changing orientations along
the circle such that the orientation is reversed upon traversing the circle
once, yielding a M\"obius strip, which is  not trivial. These are
schematically shown in Figure~\ref{fig:fiber}. 
Over a given base space vector bundles are topologically 
classified up to isomorphism. Two bundles on $M$ are isomorphic 
if on each open set $U_{\alpha}$ the vector spaces corresponding to these
are isomorphic. 

Let us next discuss the connection between the time-reversal invariant 
crystalline systems and $G$-bundles.
A Bloch wavefunction $\psi(x,k)$ is an element of the Hilbert space 
$\mathcal{H}$ which varies over the Brillouin zone $\B$
parametrized by $k$. Hence the group of endomorphisms of $\mathcal{H}$, 
which form a module over $\mathcal{H}$ varies over $\B$ too.
As the wavefunctions transform under an $\SO(3)\simeq
\SU(2)/\Z_2$, as shown before, we are led to considering 
$\SO(3)$-bundles over the torus $\B$. This corresponds to considering
wavefunctions associated to two types of band functions discussed above.
Whether or not the twisting can occur in a real system depends on the details
of the Hamiltonian. The geometric picture we consider only classifies the
different possibilities, of which there are but two, the first corresponding
to a conventional insulator while the second to a topological one.

Since the isomorphism of vector spaces on an open set is required to be
consistent with the action of the group $G$ on the fibers, a classification 
scheme for vector bundles depends on properties of $G$.
The standard procedure is to determine the homotopy class
of maps from the base space $M$ of $V$ to the 
classifying space $BG$ associated with $G$. 
Homotopy properties of $BG$ are known. In a homotopic classification
two maps belong to the same class if they can be continuously deformed
into each other.  
Fortunately we do not need to get into the details of the space $BG$ since
for vector bundles over a $D$-sphere $\sph^D$ the 
homotopy classes that need be computed depend only on the
fiber group $G$. 

We now have a well posed topological problem. The topological tool appropriate for the problem is $K$-groups . It is used to classify bundles over both spheres and tori.

In order to carry out $K$-group calculations over a base space that is not a sphere 
a number of technical issues have to be addressed. We define $K$-groups and then explain these technical matters and then return to show how these results allow us to understand the topological nature of topological insulators. However although the defining and explaining   the rules for calculating 
$K$-groups are rather technical the rules themselves 
are easy to implement.
\section{$K$-theory} \label{sec:kth}
$K$-theory may be thought of as a generalized cohomology theory
used to classify vector bundles \cite{Nash2,egh}. 
The method directly works with vector bundles and spots twists  
from the Abelian group structure introduced for adding bundles.
The Abelian group operation between bundles starts by defining addition
of bundles over the same base space simply as the usual addition of vector 
spaces. The fiber of the sum of two vector bundles on any open set of the base
space is taken to be the sum of fibers of the summands on the same open set.
A trick analogous to the definition of integers from natural numbers
then leads to the notion of subtraction of bundles. 
This yields an Abelian group, known as the $K$-group, with vector bundles 
as its elements and addition of bundles as the group operation, 
its inverse operation being the subtraction. 

Topological information is fed into the structure by stating that a pair of vector bundles are
isomorphic if they differ by the addition of a trivial bundle
of appropriate rank, the latter taken to play the role identity like
zero in the group of integers. 

A vector bundle $V^{(k)}$ of rank $k$ on a manifold 
$M$ of dimension $D$ is said to be in the \emph{stable range} if $k>D$.  
Usefulness of the $K$-group comes from
\begin{theorem}[Nash\cite{Nash2}]
If a vector bundle $V^{(k)}$ is in the stable range, then there is an
isomorphism between the vector bundles $V^{(k)}$ and $V^{m}+I^{(k-m)}$
with $m< k$, where $V^{(m)}$ is a vector bundle of rank $m$ and $I^{(k-m)}$ is
a trivial bundle of rank $(k-m)$ both defined over the base space $M$. We
write $V^{(k)} =  V^{(m)}+I^{(k-m)}$.
\end{theorem}
This means that the totality of topological information of a vector bundle of
a sufficiently large rank is effectively encoded in a vector bundle of lower
rank. Hence the following equivalence relation between vector bundles holds:
\begin{defn}
Two vector bundles $V_1$ and $V_2$ are equivalent if $V_1 + I^{\ell} =
V_2+I^{r}$, where $\ell$ and $r$ are positive integers. We write 
$V_1\sim V_2$.
\end{defn}
We have the convenient 
\begin{lemma}
For vector bundles $V_0$, $V_1$ and $V_2$, if $V_1+V_0=V_2+V_0$, 
then $V_1\sim V_2$.
\end{lemma}
The proof is based on the fact \cite{Nash2} that for a reasonable 
manifold $M$ it is always possible to find for a given vector bundle
$V_0$ on $M$,  a bundle $V_0'$, such that $V_0+V_0'=I$, $I$ being
a trivial bundle.
Thus $V_1+V_0 = V_2 +V_0$ implies $V_1+V_0 +V_0'= V_2 +V_0+V_0'$, or 
$V_1+ I =V_2+I$, that is $V_1\sim V_2$.

This equivalence relation partitions the set of vector bundles on a fixed 
base space into equivalence classes, which generalizes the isomorphism classes
in $K$-theory. The usual notion of isomorphism follows by considering vector
bundles of equal ranks. 

From now on a vector bundle will be taken to refer to a
representative of the equivalence class to which it belongs. 
We have considered addition of vector bundles and their equivalence through
the addition of a trivial bundle. Let us now discuss the subtraction of bundles. The
notion of subtraction of vector bundles parallels a particular feature of
subtraction of natural numbers. Let us consider two pairs of natural numbers
$\{v_1, v_2\}$, $\{v_1', v_2'\}$ and write their pairwise differences as the
ordered pairs, namely, $(v_1,v_2)=v_1-v_2$ and $(v_1',v_2')=v_1'-v_2'$.
If the differences are equal, that is
$(v_1,v_2)=(v_1',v_2')$,
then we can also write 
$v_1+v_2'=v_1'+v_2$,
using the usual properties of subtraction as the inverse of addition for
natural numbers, or, equivalently,
\begin{equation}
\label{eq:subtr}
v_1+v_2'+v=v_2+v_1'+v
\end{equation}
where $v$ is an arbitrary natural number. Reasoning backwards, subtraction
of two natural numbers $(s,t)$ is now defined as the ordered pair so that
$(v_1,v_2)=(v_1',v_2')$ for two pairs of natural numbers implies equation
\eq{eq:subtr}.

This notion is used to define the subtraction of vector bundles. Given a pair
of vector bundles $V_1$ and $V_2$ on $M$ their difference, also 
considered to be a vector bundle on $M$,  is defined to be an
ordered pair $(V_1,V_2)$, such that if $(V_1,V_2)=(V_1',V_2')$ for two pairs
of vector bundles, then 
\begin{equation}
V_1+V_2'+V = V_2+V_1'+V, 
\end{equation} 
for any vector bundle $V$ over $M$. The difference $(V_1,V_2)$ is 
also called a \emph{virtual bundle} and denoted  $V_1-V_2$, 
with \emph{virtual dimension} 
of $V_1-V_2$ defined as the difference between the ranks of the individual
bundles $V_1$ and $V_2$. Let us point out  that the virtual
dimension can be negative; it is zero if both the bundles have the
same rank.
The ability to add and subtract vector bundles belonging to the equivalence class of vector bundles defined gives them the structure of an Abelian group, known as the $K$-group.
The important point to note that these groups are properties of bundles,
no details regarding
their fiber space or base space structures appear in the definitions. These details show up when we want to calculate $K$-groups.
\section{Computing $K$-groups}
\label{sec:comput}
In this section we discuss how to calculate the $K$-groups of
$G$-bundles over spheres and tori. 
Let us first recall that two spaces are said to be
\emph{homeomorphic} if they are mapped to each other by a continuous 
one-to-one and onto mapping with a continuous inverse. Our first technical tool is the notion of an exact sequence of Abelian groups. This relates a collection of Abelian groups in which the image of one map between them is the kernel of the next map. Such a sequence of maps is said to be exact. These sequences summarize the way groups associated with different spaces are related. This feature will become evident when we look at specific spatial structures. We write, 
\begin{equation}
\label{esq}
0\rt A \stackrel{f}{\rt} B\stackrel{g}{\rt}  C\rt 0 
\end{equation} 
The sequence of maps is said to be \emph{exact} if the image of the group $A$ into the group $B$ under the injective mapping $f$ is mapped to the identity element of 
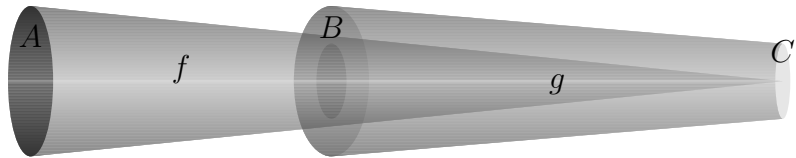
\begin{figure}[h]
\begin{center}
\begin{tikzpicture}
\fill[black!70!white] (0,0) circle (.3cm and 1cm);
\fill[black!10!white] (10,0) circle (.1cm and .5cm);
\fill[black!60!white,opacity=.25] (4,0) circle (.5cm and 1cm);
\fill[black!40!white] (4,0) circle (.2cm and .5cm);
%
\fill[top color=gray!30!black,bottom color=gray!30!black,middle
color=gray!10,shading=axis,opacity=0.25] (0,-1) -- (10,0) -- (0,1) arc
(90:270:.3cm and 1cm);   
\fill[top color=gray!50!black,bottom color=gray!40!black,middle
color=gray!10,shading=axis,opacity=0.25] (4,-1) -- (10,-.5) arc (270:90:.1cm 
and .5cm) -- (10,.5) -- (4,1) arc (90:270:.5cm and 1cm); 
\node (a) at (0,.6) {$A$}; 
\node (b) at (4,.72) {$B$}; 
\node (c) at (10,.4) {$C$}; 
\node (f) at ($(a)!.5!(b)-(0,.5)$) {$f$}; 
\node (g) at ($(b)!.5!(c)-(0,.6)$) {$g$}; 
\end{tikzpicture}
\caption{Pictorial presentation of the exact sequence \eq{esq}}
\label{fig:exact}
\end{center}
\end{figure}
the group $C$ by the surjective mapping $g$, as illustrated in 
Figure~\ref{fig:exact}. This is a manner of expressing the fact that $C$ is
same as $B$ modulo $A$, which is shrunk to a point in $C$.

$K$-groups are known for vector bundles over spheres in terms of 
homotopy groups. We
will use exact sequence arguments to relate them for bundles over tori. But to do this a conceptual problem has to be addressed. The homotopy groups for a space $X$ use maps  that all have a fixed starting point, a base point, but for building tori from spheres  joining together different spaces is required.  Each space has an associated fixed point, its base point. Hence a conceptual way of tackling this problem is required.  The intuitive idea is to identifying these base points, pictorially by smashing or collapsing  them to a single points.  This is done as follows.
 
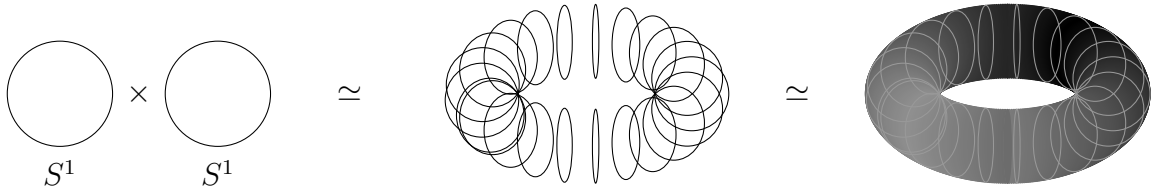
\begin{figure}[h]
\begin{center}
\begin{tikzpicture}[scale=.7]
\begin{scope}
\draw (0,0) circle (1);
\node at (1.5,0) () {$\times$};
\draw (3,0) circle (1);
\node at (0,-1.5) () {$\sph^1$};
\node at (3,-1.5) () {$\sph^1$};
\node at (5.5,0) () {$\simeq$};
\end{scope}
\begin{scope}[shift={(10,0)}]
 \foreach \x in {-153,-136,...,210} { 
    \pgfmathsetmacro\elrad{20*(cos(\x))}
    \draw (xyz polar cs:angle=\x,y radius=1,x radius=2)
        ellipse (\elrad pt and 20pt);
}
\node at (4,0) () {$\simeq$};
\end{scope}
\begin{scope}[shift={(18,0)}]
    \foreach \x in {90,89,...,-90} { 
    \pgfmathsetmacro\elrad{20*max(cos(\x),.1)}
    \pgfmathsetmacro\ltint{0.64706*abs(\x-45)/180}
    \pgfmathsetmacro\rtint{0.564706*(1-abs(\x+45)/180)}
    \definecolor{currentcolor}{rgb}{\ltint, \ltint, \ltint}
    \draw[color=currentcolor,fill=currentcolor]
        (xyz polar cs:angle=\x,y radius=1,x radius=2)
        ellipse (\elrad pt and 20pt);
    \definecolor{currentcolor}{rgb}{\rtint,  \rtint, \rtint}
    \draw[color=currentcolor,fill=currentcolor]
        (xyz polar cs:angle=180-\x,radius=1,x radius=2)
        ellipse (\elrad pt and 20pt);
    }
 \foreach \x in {-153,-136,...,210} { 
    \pgfmathsetmacro\elrad{20*(cos(\x))}
    \draw [color=white!60!black] (xyz polar cs:angle=\x,y radius=1,x radius=2)
        ellipse (\elrad pt and 20pt);
}
\end{scope}
\end{tikzpicture}
\end{center}
\caption{Cartesian product of circles: two-torus}
\label{fig:Cartesian}
\end{figure}

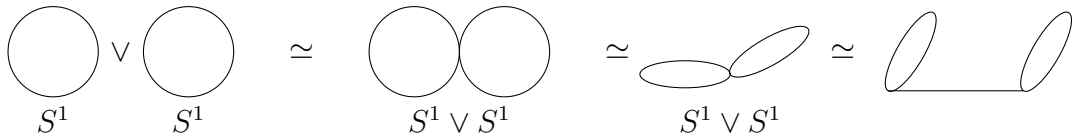
\begin{figure}[b]
\begin{center}
\begin{tikzpicture}[scale=.6]
\begin{scope}
\draw (0,0) circle (1);
\node at (1.5,0) () {$\vee$};
\draw (3,0) circle (1);
\node at (0,-1.5) () {$\sph^1$};
\node at (3,-1.5) () {$\sph^1$};
\node at (5.5,0) () {$\simeq$};
\end{scope}
\begin{scope}[shift={(8,0)}]
\draw (0,0) circle (1);
\draw (2,0) circle (1);
\node at (1,-1.5) () {$\sph^1\vee \sph^1$};
\node at (4.5,0) () {$\simeq$};
\end{scope}
\begin{scope}[shift={(14,-.5)}]
    \pgfmathsetmacro\ellsftcos{1+1*cos(30)}
    \pgfmathsetmacro\ellsftsin{1*sin(30)}
    \draw (0,0) ellipse (1 and .3);
    \draw [rotate around={30:(\ellsftcos,\ellsftsin)}]  (\ellsftcos
,\ellsftsin ) ellipse (1 and .3);
\node at (1,-1) () {$\sph^1\vee \sph^1$};
\end{scope}
\begin{scope}[shift={(16,0)}]
\node at (1.5,0) () {$\simeq$};
    \pgfmathsetmacro\rotangle{60}
    \pgfmathsetmacro\elbotlx{3-cos(\rotangle)}
    \pgfmathsetmacro\elbotly{-sin(\rotangle)}
    \pgfmathsetmacro\elbotrx{6-cos(\rotangle)}
    \pgfmathsetmacro\elbotry{-sin(\rotangle)}
    \draw  [rotate around={\rotangle:(3,0)}] (3,0) ellipse (1 and .3);
    \draw  [rotate around={\rotangle:(6,0)}] (6,0) ellipse (1 and .3);
\draw (\elbotlx , \elbotly) -- (\elbotrx ,\elbotry);
\end{scope}
\end{tikzpicture}
\end{center}
\caption{Wedge sum of two circles}
\label{fig:wedge}
\end{figure}

Consider $X$ and $Y$ that be two (topological) spaces with base-points, 
that is each one has one point marked as distinguished, 
which we denote by $x_0$ and $y_0$, respectively. Then 
the \emph{Cartesian product} $X\times Y$ of $X$ and $Y$ is defined to be the 
set of ordered pairs of points from $X$ and $Y$, namely
\begin{equation}
X\times Y = \{(x,y)|x\in X, y\in Y\}.
\end{equation} 
For example,  a two-torus is (homeomorphic to) the Cartesian product to two
circles, $\torus{2}\simeq \sph^1\times \sph^2$, with  
$X=\sph^1$ and $Y=\sph^1$, as depicted in Figure~\ref{fig:Cartesian}.
The \emph{wedge sum} $X\vee Y$ is the disjoint union of $X$ and $Y$ but with
the base-points identified. This space is essential for fixing a point in a torus. That is 
\begin{equation}
X\vee Y = X\sqcup Y/(x_0\sim y_0). 
\end{equation} 
For example, the wedge sum of two circles is (homeomorphic to) a figure eight
as indicated in Figure~\ref{fig:wedge}.
%
The \emph{smash product} $X\wedge Y$ of $X$ and $Y$ is defined as 
their Cartesian product with the subsets $(x_0,Y)=\{(x_0,y)|y\in Y\}$ 
and $(X,y_0)=\{(x,y_0)|x\in X\}$ identified. Thus
\begin{equation}
X\wedge Y = X\times Y/\big((x_0,y)\sim (x,y_0)\big),\quad \forall x\in X,
\forall y\in Y.
\end{equation} 
The subsets $X$ and $Y$ of $X\times Y$ can be looked upon as the 
subspaces $X\times\{y_0\}$ and  $\{x_0\}\times Y$, respectively, intersecting
at $x_0,y_0$, which is taken to be the base-point of $X\times Y$. Thus their
union can be identified with the wedge sum $X\vee Y$. Hence the smash product
can be expressed as a quotient
\begin{equation}
X \wedge Y = (X \times Y) / (X \vee Y),
\end{equation} 
which can also be described as the sequence of mappings
\begin{equation}
\label{seq:smash}
 0 \rt X \vee Y \rt X \times Y \rt X \wedge Y \rt 0,
\end{equation}
where the first map is injective taking a point, denoted $0$, 
to the basepoint of $X\vee Y$, while the last
map is surjective collapsing the entire space $X\wedge Y$ to a point, as
depicted in Figure~\ref{fig:exact}.

\begin{figure}[h]
\begin{center}
\begin{tikzpicture}[scale=.7]
\begin{scope}
\draw (0,0) circle (1);
\node at (1.5,0) () {$\times$};
\draw (3,0) circle (1);
\node at (0,-1.5) () {$\sph^1$};
\node at (3,-1.5) () {$\sph^1$};
\node at (7.5,0) () {$\simeq$};
\end{scope}
\begin{scope}[shift={(10,-1)}]
 \foreach \x in {10,30,...,170} { 
    \pgfmathsetmacro\elrad{50*(sin(\x))}
    \draw (\x pt,  \elrad pt) ellipse (5pt and \elrad pt );
}
\draw (10 pt,0) to (170 pt,0);
\end{scope}
\end{tikzpicture}
\end{center}
\caption{Alternative view of Cartesian product of two circles}
\label{fig:wedg2}
\end{figure}
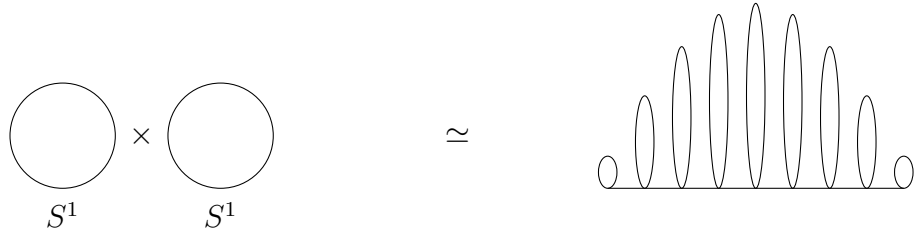

\begin{figure}[b]
\begin{center}
\begin{tikzpicture}[scale=.6]
\begin{scope}
\draw (0,3) circle (1);
\node at (1.5,3) () {$\times$};
\draw (3,3) circle (1);
\node at (0,-1.5) () {$\sph^1$};
\node at (3,-1.5) () {$\sph^1$};
\draw (-2,1.5) -- (5,1.5);
\draw (0,0) circle (1);
\node at (1.5,0) () {$\vee$};
\draw (3,0) circle (1);
\node at (0,-1.5) () {$\sph^1$};
\node at (3,-1.5) () {$\sph^1$};
\end{scope}
\begin{scope}[shift={(7,2)}]
 \foreach \x in {10,30,...,170} { 
    \pgfmathsetmacro\elrad{50*(sin(\x))}
    \draw (\x pt,  \elrad pt) ellipse (5pt and \elrad pt );
}
\draw (10 pt,0) to (170 pt,0);
\draw (0 pt,-.5) to (180 pt,-.5);
\node at (-25 pt,-.5) () {$\simeq$};
\node at (205 pt,-.5) () {$\simeq$};
    \pgfmathsetmacro\rotangle{60}
    \pgfmathsetmacro\elbotlx{1.7-cos(\rotangle)}
    \pgfmathsetmacro\elbotly{-2-sin(\rotangle)}
    \pgfmathsetmacro\elbotrx{5-cos(\rotangle)}
    \pgfmathsetmacro\elbotry{-2-sin(\rotangle)}
    \draw  [rotate around={\rotangle:(1.7,-2)}] (1.7,-2) ellipse (1 and .3);
    \draw  [rotate around={\rotangle:(5,-2)}] (5,-2) ellipse (1 and .3);
\draw (\elbotlx , \elbotly) -- (\elbotrx ,\elbotry);
\end{scope}
%
\begin{scope}[shift={(17,0.5)}]
 \foreach \x in {10,30,...,170} { 
    \pgfmathsetmacro\elrad{50*(sin(\x))}
    \pgfmathsetmacro\elx{\elrad*(cos(\x))}
    \pgfmathsetmacro\ely{\elrad*(sin(\x))}
    \draw [rotate around={\x:(\elx pt,\ely pt)}] (\elx pt, \ely pt) ellipse (\elrad pt and 5pt );
}
\node at (2.8,1) () {$\simeq$};
\end{scope}
\begin{scope}[shift={(22.5,1.3)}]
\def\R{1.8} 
\def\angEl{35} 
\filldraw[ball color=white] (0,0) circle (\R);
\foreach \t in {-5,-35,...,-175} { \DrawLongitudeCircle[\R]{\t} }
\end{scope}
\end{tikzpicture}
\end{center}
\caption{Smash product of two circles as a quotient}
\label{fig:spher}
\end{figure}
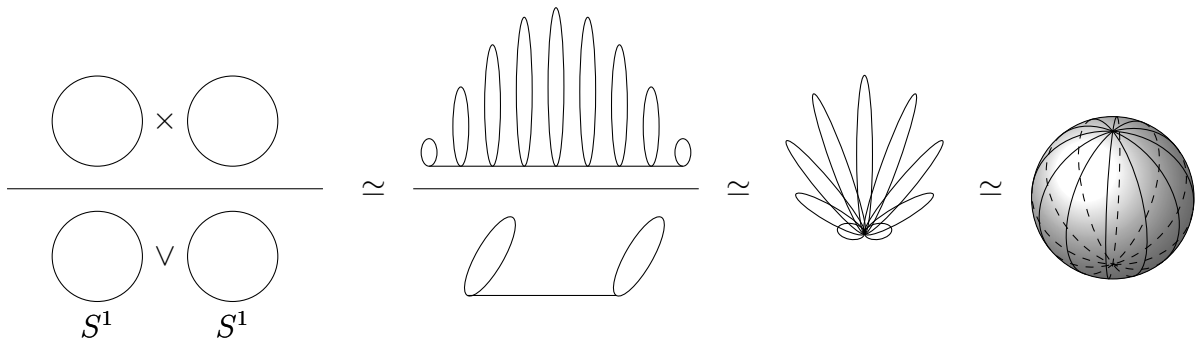

The smash product of two circles is (homeomorphic to) a sphere, $\sph^2$. 
To see this we first ``open up" one circle in each of $\sph^1\times \sph^1$ and
$\sph^1\wedge \sph^1$ to an interval, ignoring the identification of endpoints.  
For the Cartesian product $\sph^1\times \sph^1$ this is depicted in 
Figure~\ref{fig:wedg2}, where the radii of the circles along the interval
is varied, decreasing from the center towards the
end on both sides, without affecting the topology of the configuration. 
Upon identification of the end points it becomes the torus of 
Figure~\ref{fig:Cartesian}. 
Similarly, the figure eight wedge sum
$\sph^1\vee \sph^1$ is opened up as indicated in the
last diagram of Figure~\ref{fig:wedge} with the two circles placed at the end
points of an interval, which is contractible to a point. The figure eight is
recovered upon identification of the end points of the interval. 
We can now describe the quotient $\sph^1\wedge \sph^1$ pictorially as
the quotient of these two pictures, 
shrinking the interval in Figure~\ref{fig:wedg2} with the circles at its
end points brought together to a point due to the identification by the
configuration in the last diagram of Figure~\ref{fig:wedge}. This yields a
sphere $\sph^2$ as indicated in Figure~\ref{fig:spher}.
Coming back to the calculation of $K$-groups, we can progress by considering a sequence of spaces and associating with them an associated an exact sequence of the $K$-groups of the spaces in reverse order. Thus, an exact sequence of $K$-groups 
\begin{equation}
0 \rt K(X \wedge Y) \rt K(X \times Y) \rt K(K \vee Y) \rt 0
\end{equation}
is associated to the  sequence \eq{seq:smash}. 
Here we restrict our attention to the \emph{reduced} $K$-group, which we continue
to write as  $K$. This step essentially leaves out $K$-group terms that come from points.
We now state our two operational result. First
\begin{equation}
 K(X \times Y)=K(X \wedge Y) + K(X \vee Y).
\end{equation}
Second, 
\begin{equation}
 K(X \vee Y)=K(X) + K(Y).
\end{equation}
Combining these we have the 
\begin{equation}
\label{eq:key}
 K(X \times Y)=K(X \wedge Y)+K(X) + K(Y).
\end{equation}
This equation is the key computational result. It relates the product of two spaces  $(X,Y)$to a sum of three terms, namely two terms are simply the groups associated with each of the spaces $X$ and $Y$ while the other is one space constructed from the two spaces $(X,Y)$.
We note  that we had shown earlier that $\sph^1\wedge \sph^1=\sph^2$. This is a key result.

We now state the important result that the reduced $K$-group of a $G$-bundle on a $D$-dimensional sphere is given by 
\begin{equation}
\label{k:pi}
K_G(\sph^D)=\pi_{D-1}(G),
\end{equation}
where $\pi_{n}(G)$ denotes the $n$-th homotopy group of $G$. Thus if the $K$-group for a torus is related to those over spheres we will have the result we want.

Let us sketch the proof of this result.  A sphere 
$\sph^D$ can be represented topologically by its cover with two 
contractible spaces $U_1$ and $U_2$. These spaces
overlap to form the topological space $\sph^{D-1}$. As $U_1$ and $U_2$ 
are contractible spaces homeomorphic to $\R^D$, bundles on them 
are trivial, that is direct products, $U_1\times F$ and $U_2\times F$,
respectively, where $F$ is the fiber space on which the group structure group
$G$ acts. The nature of bundles on $\sph^D$ thus depends on the
way of gluing the trivial bundles on $U_1$ and $U_2$ together. This in turn
depends on the homotopy properties of the map from the intersection
of $U_1$ and $U_2$ to $G$, namely, $\sph^{D-1}\simeq U_1\cap U_2\to G$. 
But this is just $\pi_{D-1}(G)$, by definition. 
\section{$K$-groups for time-reversal invariant systems}
\label{sec:topinsul}
We can now calculate the reduced $K$-groups of $G=\SO(3)$-bundles on
the tori $\B$ for a spin-orbit, time-reversal invariant  
system. For two- and three-dimensional systems $\B=\torus{2}$ and 
$\B=\torus{3}$, respectively. 

In the two-dimensional case by the identification of the Cartesian product 
of two circles with a two-torus and the smash product of two circles with 
a two sphere, we obtain, 
\begin{equation}
\label{eq:keysph}
\begin{split}
K_G(\torus{2}) &= K_G(\sph^1\times \sph^1)\\
&=K_G(\sph^1\wedge \sph^1)+K_G(\sph^1) + K_G(\sph^1)\\
&=K_G(\sph^2) + 2K_G(\sph^1),
\end{split}
\end{equation}
where the notation $K(X)+K(X)=2K(X)$ is used for $K$-groups in analogy with
the natural numbers.
Next, using  \eq{k:pi} equation \eq{eq:keysph} yields
\begin{equation} 
K_G(\torus{2}) = \pi_1(G) + 2\pi_0(G),
\end{equation} 
where $G=\SO(3)$. 
Ignoring the last term in reduced $K$-theory we then obtain 
\begin{equation} 
K_{\SO(3)}(\torus{2}) = \pi_1\big(\SO(3)\big) = \Z_2.
\end{equation} 
The two classes in the $K$-group represent  the two types of bundles
corresponding to the osculating and intersecting band functions discussed in
section~\ref{sec:topology}.

For the  three-dimensional system we compute the reduced $K$-group of
$\SO(3)$-bundles on $\torus{3}=\sph^1\times \sph^1\times \sph^1$. Defining
$Y=\sph^1\times \sph^1\simeq\torus{2}$ we use \eq{eq:key} to obtain 
\begin{equation}
\begin{split}
K_G(\torus{3})&=K_G(\sph^1 \times Y)\\
&=K_G(\sph^1 \wedge Y) + K_G(\sph^1) + K_G(Y)\\
&=K_G(\sph^1 \wedge Y) + \pi_0(G) + K_G(\torus{2})\\
&=K_G(\sph^1 \wedge Y) + \pi_1(G),
\end{split}
\end{equation}
where we have used the results for the two-torus and ignored $\pi_0$.
In order to calculate $K_G(\sph^1 \wedge Y)$ we need the
\begin{theorem}[James\cite{James2}]
The $r$-th Betti number $b_r$ of $Y$ contributes a $K_G(\sph^{r+1})$ to
the $K$-group of the wedge sum $K_G(\sph^1\wedge Y)$.
\end{theorem}
This arises from the fact that the $r$-th Betti number counts the number 
of $r$-th 
homology cycles of $Y$ which are $r$-spheres. These contribute to the
$K$-group in one higher dimension due to the extra $\sph^1$.
Betti numbers of  $Y\simeq\torus{2}$ are $b_0=1,b_1=2,b_2=1$. 
Hence 
\begin{equation} 
K_G(\sph^1\wedge Y)= K_G(\sph^1) + 2K_G(\sph^2) + K_G(\sph^3).
\end{equation} 
We thus have, up to $\pi_0$ of circles and spheres,
\begin{equation}
K_G(\torus{3}) =3\Z_2 + K_G(\sph^3)
\end{equation}
in reduced $K$-theory. The last term requires special treatment not being in
the stable range. 
We need the following
\begin{theorem}[{James \& Thomas\cite[Th.\ 1.6]{James1}}]
The map $[T^3,B\SO(3)] \rt [T^3,BSO] = K(T^3)$ is injective, and under  this map the elements of $[T^3,B\SO(3)]$ correspond to the subgroup of $K(T^3)$ with vanishing third Steifel Whitney class.
\end{theorem}
Here $T^3=\sph^1 \times \sph^1 \times \sph^1$ and 
the Steifel Whitney class $H^3(\sph^3,\Z_2)=\Z_2$ is non-zero.  Hence 
$K(T^3)$ is the trivial group. 
Thus 
\begin{equation}
K_{\SO(3)}(\torus{3}) = 3\Z_2,
\end{equation}
which agrees with previous results \cite{fu,MB,R}.

In section \ref{sec:topology} we proved the existence of Dirac cones 
as a consequence of gap-less states appearing as band functions intersect.
Let us briefly discuss how the appearance of gap-less states follow 
from $K$-theory analysis. 

Here we use the fact that spin-orbit systems can be viewed as Dirac equations
in the presence of $\frac{\SU(2)}{Z_2}$ gauge fields. Recall this follows
from the fact that an orbital interaction has the structure $\sigma.g(x)\vec{L}$ 
where $\vec{L}$ is a generator of the rotation group $\SO(3)=\frac{\SU(2)}{Z_2}$ .
We also showed how the group $\SO(3)$ appears at Kramer points of time-reversal invariance. The next step is to use two powerful result of topology, the index theorem\cite{Nash2}.
This theorem relates the zeros of Dirac operators to a topological property
of the system. The index of the
Dirac operator $D$, which acts on smooth sections of the 
$\SU(2)/\Z_{2}$-bundle on  the torus $\B$, is defined as 
\begin{equation}
\begin{split}
\ind{D} &= \ch{\ker{D} - \coker{D}}\\
&= \ch{\ker{D}} - \ch{\coker{D}},
\end{split}
\end{equation} 
and the result that the Chern character is a topological map from the $K$-group of a space to its rational cohomology, namely,
\begin{equation}
\operatorname{ch}: K(X) \to H^{\bullet}(X,\mathbf{Q})
\end{equation} 
Thus we have a map of $\ker{D} - \coker{D}$ to 
$K_G(\B)$ to $H^{\bullet}(\B,Q)$. We have already calculated the $K$-group and shown that it  is non-zero for surface points. Thus  either the kernel or its dual, the cokernel, of $D$ is non-void, implying the existence of zero modes, 
or gap-less states, at some point of $\B$ where there is conductance.
\section{Conclusion}
\label{sec:concl}
The procedure of calculating $K$-groups described works for
 bundles over base manifolds that are tori. We illustrated the method by explicitly 
calculating the $K$-groups for $\SO(3)$-bundles over two- and three-dimensional tori
and explained why they explain the topological origin of gap-less surface points on a topological insulator. Two steps were involved. In the first  relevant $K$-groups were
calculated and it was shown that for the bulk the $K$ groups were zero but
for special surface points, Kramer points, they were non zero. 
In the next step the non vanishing of a $K$-group at surface points was shown to
imply the existence of gap-less surface states by using the 
index theorem, a tool for finding gap-less (zero modes) for the Dirac operator
and the link between it and $K$ groups. Thus the topological origin
of gap-less states was established. In the second step the strong spin-orbit interactions
and the time-reversal invariant nature of the topological insulator played a 
key role as they justify the use of a Dirac operator at Kramer points.

The calculation of $K$-groups over tori was carried out by showing how 
they could be reduced to calculations over spheres with the help of formula 
that related  $K$-group calculations  for spaces 
$S\wedge Y$ to ones over spheres\cite{James1,James2}.
Our results agree with previous intuitive results.

Although our calculations  used a number of tools and ideas of topology that
are not widely known and thus required explaining, the calculation themselves involved, as seen a small number of simple steps. Theorems of James were used to relate a given base space to spheres. Then $K$-groups on a sphere were related to a homotopy group and finally  for spaces of the form $S\wedge Y$ an algebraic  procedure for calculating $K$-groups was described. 

In the approach the existence of conduction points was crucially 
 related to special Kramer points of a time-reversal invariant system
and the non vanishing of the index of the Dirac operator. The Dirac operator, a relativistic
operator, appears in this non relativistic framework because, as we saw, the system has
 strong spin-orbit interactions and time-reversal symmetry.

Let us comment on the importance of time-reversal invariance in the calculation. If the structure group of the bundle was been $\SU(k)$ for any $k$, \ie had we were dealing with complex bundles, then the $K$-groups would have been trivial \cite{Nash2}. There would not be any gap-less states. But time-reversal symmetry breaks $\SU(2)$ to $\SO(3)$  and for these groups the surface $K$-groups are non trivial.

Let us explain this feature of the calculation. The base space of the bundle of interest  is the Brillouin zone. This is a compact $T^3$ surface  in momentum space and reflects space periodicity that is present for the system. The Brillouin zone has no boundary. However in momentum space that the effect of time-reversal symmetry gives rise to Kramer degeneracy, namely the degeneracy of spin-up spin-down states at Kramer points. Identifying these  points in momentum space generates a topological $\Z_2$ twist. In this picture the surface of the topological insulator in space is reflected by surface sections , 
$T^2=\sph^1\times \sph^1$ parts of the Brillouin zone three torus. There are three independent surface sections in $T^3$ that give the three $\Z_2$ contributions
that were found. The  $T^2$ spaces, are closed  surface sections
of $T^3$. However they do represent the surface effects of the topological insulator because the relevant bulk topological insulator surface of $T^3$ are  three $T^2$ surfaces. It is these surfaces that have  non zero $K$-groups and gap-less states.  Thus our calculations in momentum space do show that only the surface of the topological insulator can have gap-less states.

The $K$-groups defined have an additional algebraic property. Their elements can be added to form an algebraic ring. The group property of a set  is usually represented as a multiplication and a ring is a set with elements with rules for both multiplication and addition. This ring  property can be used to extend the results obtained for a single gap-less point, a Kramer point with Dirac operator, a Dirac point, to the case where there are multiple Dirac points. In this case  the relevant $K$-group is the sum of the $K$-groups of the factors, one for each Dirac point. This implies triviality of $K$-groups for an even number of Dirac points and non-triviality for an odd number as the sum of an even number of $\Z_2$'s is trivial but is not for an odd number of $\Z_2$'s.  

Finally we note that since  all condensed matter systems with periodicity have Brilloun zones  the method of calculating their topological properties using $K$ groups described here are applicable. 
\subsection*{Acknowledgement}
SS acknowledges the hospitality of Theoretical Physics Department,
IACS during completion of this work which is based on a set of
lectures delivered at IACS, and thanks C. Nash for sending him
a copy of  \cite{James1}. KR thanks P. Majumdar,
A. Mukherjee and K. Sengupta for fruitful discussions.

\end{document}